\documentclass[twoside,spanish,english,prl,twocolumn,showpacs]{revtex4}
\usepackage{mathpazo}
\usepackage{helvet}
\usepackage{beramono}
\usepackage[OT1]{fontenc}
\usepackage[latin9]{inputenc}
\usepackage[letterpaper]{geometry}
\usepackage{babel}
\addto\shorthandsspanish{\spanishdeactivate{~<>}}

\usepackage{amsmath}
\usepackage{amssymb}
\usepackage{graphicx}
\usepackage{esint}
\usepackage[colorlinks=false]{hyperref}

\makeatletter
\@ifundefined{textcolor}{}
{%
 \definecolor{BLACK}{gray}{0}
 \definecolor{WHITE}{gray}{1}
 \definecolor{RED}{rgb}{1,0,0}
 \definecolor{GREEN}{rgb}{0,1,0}
 \definecolor{BLUE}{rgb}{0,0,1}
 \definecolor{CYAN}{cmyk}{1,0,0,0}
 \definecolor{MAGENTA}{cmyk}{0,1,0,0}
 \definecolor{YELLOW}{cmyk}{0,0,1,0}
}

\usepackage{microtype}

\makeatother

\begin{document}

\title{Error analysis of free probability approximations to the density
of states of disordered systems}

\author{Jiahao Chen}

\email{jiahao@mit.edu}

\selectlanguage{english}%

\affiliation{Department of Chemistry, Massachusetts Institute of Technology, 77
Massachusetts Avenue, Cambridge, Massachusetts 02139, USA}

\author{Eric Hontz}

\email{ehontz@mit.edu}

\selectlanguage{english}%

\affiliation{Department of Chemistry, Massachusetts Institute of Technology, 77
Massachusetts Avenue, Cambridge, Massachusetts 02139, USA}

\author{Jeremy Moix}

\email{jmoix@mit.edu}

\selectlanguage{english}%

\affiliation{Department of Chemistry, Massachusetts Institute of Technology, 77
Massachusetts Avenue, Cambridge, Massachusetts 02139, USA}

\selectlanguage{spanish}%

\author{Alberto Su\'arez}

\email{alberto.suarez@uam.es}

\selectlanguage{spanish}%

\affiliation{Departamento de Ingenier\'{\i}a Inform\'atica, Escuela Polit\'ecnica
Superior, Universidad Aut\'onoma de Madrid, Ciudad Universitaria
de Cantoblanco, Calle Francisco Tom\'as y Valiente, 11 E-28049 Madrid,
Spain}

\selectlanguage{english}%

\author{Ramis Movassagh}

\email{ramis@mit.edu}

\selectlanguage{english}%

\affiliation{Department of Mathematics, Massachusetts Institute of Technology,
77 Massachusetts Avenue, Cambridge, Massachusetts 02139, USA}

\author{Matthew Welborn}

\email{welborn@mit.edu}

\selectlanguage{english}%

\affiliation{Department of Chemistry, Massachusetts Institute of Technology, 77
Massachusetts Avenue, Cambridge, Massachusetts 02139, USA}

\author{Alan Edelman}

\email{edelman@math.mit.edu}

\selectlanguage{english}%

\affiliation{Department of Mathematics, Massachusetts Institute of Technology,
77 Massachusetts Avenue, Cambridge, Massachusetts 02139, USA}

\author{Troy Van Voorhis}

\email{tvan@mit.edu}

\selectlanguage{english}%

\affiliation{Department of Chemistry, Massachusetts Institute of Technology, 77
Massachusetts Avenue, Cambridge, Massachusetts 02139, USA}
\begin{abstract}
Theoretical studies of localization, anomalous diffusion and ergodicity
breaking require solving the electronic structure of disordered systems.
We use free probability to approximate the ensemble-averaged density
of states without exact diagonalization. We present an error analysis
that quantifies the accuracy using a generalized moment expansion,
allowing us to distinguish between different approximations. We identify
an approximation that is accurate to the eighth moment across all
noise strengths, and contrast this with the perturbation theory and
isotropic entanglement theory.
\end{abstract}


\maketitle
Disordered materials have long been of interest for their unique physics
such as localization~\cite{Thouless1974,Evers:2008p706}, anomalous
diffusion~\cite{Bouchaud1990,Shlesinger1993} and ergodicity breaking~\cite{Palmer1982}.
Their properties have been exploited for applications as diverse as
quantum dots~\cite{Barkai2004,Stefani2009}, magnetic nanostructures~\cite{Hernando1999},
disordered metals~\cite{Dyre2000,Dugdale2005}, and bulk heterojunction
photovoltaics~\cite{Peet2009,Difley2010,Yost2011}. Despite this,
theoretical studies are complicated by the need to calculate the electronic
structure of the respective systems in the presence of random atomic
nuclear positions. Conventional electronic structure theories can
only be used in conjunction with explicit sampling of thermodynamically
accessible regions of phase space, which make such calculations enormously
more expensive than usual single-point calculations~\cite{Kollman1993}.

Alternatively, ensemble-averaged quantities may be computed or approximated
using random matrix theory. In particular, techniques from free probability
theory allow the computation of eigenvalues of sums of certain matrices
without rediagonalizing the matrix sums~\cite{Voiculescu1991}. While
this has been proposed as a tool applicable to general random matrices~\cite{Biane1998}
and has been used for similar purposes in quantum chromodynamics~\cite{Zee1996a},
we are not aware of any quantification of the accuracy of this approximation
in practice. We provide herein a general framework for quantitatively
estimating the error in such situations. We find that this allows
us to understand the relative performances of various approximations,
and furthermore characterize the degree of accuracy systematically
in terms of discrepancies in particular moments of the probability
distribution functions (PDFs).

\paragraph{Quantifying the error in approximating a PDF using free probability.---}

We propose to quantify the deviation between two PDFs using moment
expansions. Such expansions are widely used to describe corrections
to the central limit theorem and deviations from normality, and are
often applied in the form of Gram--Charlier and Edgeworth series~\cite{Stuart1994,Blinnikov1998a}.
Similarly, deviations from non-Gaussian reference PDFs can be quantified
using generalized moment expansions. For two PDFs $w\left(\xi\right)$
and $\tilde{w}\left(\xi\right)$ with finite cumulants $\kappa_{1},\kappa_{2},\dots$
and $\tilde{\kappa}_{1},\tilde{\kappa}_{2},\dots$, and moments $\mu_{1},\mu_{2},\dots$
and $\tilde{\mu}_{1},\tilde{\mu}_{2},\dots$ respectively, we can
define a formal differential operator which transforms $\tilde{w}$
into $w$ and is given by~\cite{Wallace1958,Stuart1994} 
\begin{equation}
w\left(\xi\right)=\exp\left[\sum_{n=1}^{\infty}\frac{\kappa_{n}-\tilde{\kappa}_{n}}{n!}\left(-\frac{d}{d\xi}\right)^{n}\right]\tilde{w}\left(\xi\right).\label{eq:formal-expansion}
\end{equation}
This operator is parameterized completely by the cumulants of both
distributions.

The first $k$ for which the cumulants $\kappa_{k}$ and $\tilde{\kappa}_{k}$
differ then allows us to define a degree to which the approximation
$w\approx\tilde{w}$ is valid. Expanding the exponential and using
the well-known relationship between cumulants and moments allows us
to state that if the first $k-1$ cumulants agree, but the $k$th
cumulants differ, that this is equivalent to specifying that
\begin{equation}
w\left(\xi\right)=\tilde{w}\left(\xi\right)+\frac{\mu_{k}-\tilde{\mu}_{k}}{k!}\left(-1\right)^{k}\tilde{w}^{\left(k\right)}\left(\xi\right)+O\left(\tilde{w}^{\left(k+1\right)}\right).\label{eq:error-term}
\end{equation}

At this point we make no claim on the convergence of the series defined
by the expansion of (\ref{eq:formal-expansion}), but use it as a
justification for calculating the error term defined in (\ref{eq:error-term}).
We will examine this claim later.

\paragraph{The free convolution.---}

We now take the PDFs to be DOSs of random matrices. For a random matrix
$Z$, the DOS is defined in terms of the eigenvalues $\left\{ \lambda_{n}^{\left(m\right)}\right\} $
of the $M$ samples $Z_{1},\ldots,Z_{m},\dots,Z_{M}$ according to
\begin{equation}
\rho^{\left(Z\right)}\left(\xi\right)=\lim_{M\rightarrow\infty}\frac{1}{M}\sum_{m=1}^{M}\frac{1}{N}\sum_{n=1}^{N}\delta\left(\xi-\lambda_{n}^{\left(m\right)}\right).
\end{equation}
The central idea using free probability to calculate approximate DOSs
is to split the Hamiltonian $H=A+B$ into two matrices $A$ and $B$
whose DOSs, $\rho^{\left(A\right)}$ and $\rho^{\left(B\right)}$
respectively, can be determined easily. The eigenvalues of the sum
is in general not the sum of the eigenvalues; instead, we approximate
the exact DOS with the free convolution $A\boxplus B$, i.e.~$\rho^{\left(H\right)}\approx\rho^{\left(A\boxplus B\right)}$,
a particular kind of ``sum'' which can be calculated without exact
diagonalization of $H$. The moment expansion presented above quantifies
the error of this approximation in terms of the onset of discrepancies
between the $k$th moment of the exact DOS, $\mu_{k}^{\left(H\right)}$,
and that for the free approximant $\mu_{k}^{\left(A\boxplus B\right)}$.
By definition, the exact moments are 
\begin{equation}
\mu_{k}^{\left(H\right)}=\mu_{k}^{\left(A+B\right)}=\left\langle \left(A+B\right)^{k}\right\rangle ,
\end{equation}
where $\left\langle Z\right\rangle =\mathbb{E}\mbox{ Tr }\left(Z\right)/N$
denotes the normalized expected trace (NET) of the $N\times N$ matrix
$Z$. The $k$th moment can be expanded using the (noncommutative)
binomial expansion of $\left(A+B\right)^{k}$; each resulting term
will have the form of a joint moment $\left\langle A^{n_{1}}B^{m_{1}}\cdots A^{n_{r}}B^{n_{r}}\right\rangle $
with each exponent $n_{s},m_{s}$ being a positive integer such that
$\sum_{s=1}^{r}\left(n_{s}+m_{s}\right)=k$. The free convolution
$\tilde{\mu}_{k}$ is defined similarly, except that $A$ and $B$
are assumed to be freely independent, and therefore that each term
must obey, by definition~\cite{Nica2006a}, relations of the form

\begin{subequations}
\begin{align}
0 & =\left\langle \Pi_{s=1}^{r}\left(A^{n_{s}}-\left\langle A^{n_{s}}\right\rangle \right)\left(B^{m_{s}}-\left\langle B^{m_{s}}\right\rangle \right)\right\rangle \label{eq:centered-joint-moment}\\
 & =\left\langle \Pi_{s=1}^{r}A^{n_{s}}B^{m_{s}}\right\rangle +\mbox{lower order terms},\label{eq:free-indep-matrices}
\end{align}
\end{subequations}where the degree $k$ is the sum of exponents $n_{s}$,
$m_{s}$ and the second equality is formed by expanding the first
line using linearity of the NET\@. For $k\le3$, this is identical
to the statement of (classical) independence~\cite{Nica2006a}.

For arbitrary matrices $A$ and $B$, we can construct a free approximant
\begin{equation}
Z=A+Q^{-1}BQ,\label{eq:free-approximant}
\end{equation}
where $Q$ is a $N\times N$ random matrix of Haar measure. For real
symmetric $A$ and $B$ it is sufficient to consider orthogonal matrices
$Q$, which can be generated from the $QR$ decomposition of a Gaussian
orthogonal matrix~\cite{Diaconis2005}. (This can be generalized
readily to unitary and symplectic matrices for complex and quaternionic
Hamiltonians respectively.) The effect of the similarity transformation
$Q^{-1}\cdot Q$ is to apply a random rotation to the basis of $B$
with respect to $A$, and so in the $N\rightarrow\infty$ limit of
large matrices, the density of states $\rho^{\left(Z\right)}$ converges
to the free convolution $A\mbox{\ensuremath{\boxplus}}B$~\cite{Voiculescu1991,Voiculescu1994},
i.e. that $\mu_{k}^{\left(Z\right)}=\mu_{k}^{\left(A\boxplus B\right)}$
for all $k$. This provides a numerical sampling method for calculating
the moments of the free convolution.

Testing for $\mu_{k}^{\left(A+B\right)}\ne\mu_{k}^{\left(A\boxplus B\right)}$
then reduces to testing whether each centered joint moment of the
form in (\ref{eq:centered-joint-moment}) is statistically nonzero.
The cyclic permutation invariance of the NET means that the enumeration
of all the centered joint moments of degree $k$ is equivalent to
the combinatorial problem of generating all binary necklaces of length
$k$, for which efficient algorithms exist~\cite{Sawada2001}.

The procedure we have described above allows us to ascribe a degree
$k$ to the approximation $\rho^{\left(H\right)}\approx\rho^{\left(A\boxplus B\right)}$
given the splitting $H=A+B$. For each positive integer $n$, we generate
all unique centered joint moments of degree $n$, and test if they
are statistically nonzero. The lowest such $n$ for which there exists
at least one such term is the degree of approximation $k$. This is
the main result of our paper. We expect that $k\ge4$ in most situations,
as the first three moments of the exact and free PDFs match under
very general conditions~\cite{Movassagh2010}. However, we have found
examples, as described in the next section, where it is possible to
do considerably better than degree 4.

\paragraph{Decomposition of the Anderson Hamiltonian.---}

As an illustration of the general method, we focus on Hamiltonians
of the form
\begin{equation}
H=\left(\begin{array}{cccc}
h_{1} & J\\
J & h_{2} & \ddots\\
 & \ddots & \ddots & J\\
 &  & J & h_{N}
\end{array}\right),\label{eq:anderson}
\end{equation}
where $J$ is constant and the diagonal elements $h_{i}$ are identically
and independently distributed (iid) random variables with probability
density function (PDF) $p_{h}\left(\xi\right)$. This is a real, symmetric
tridiagonal matrix with circulant (periodic) boundary conditions on
a one-dimensional chain. Unless otherwise stated, we assume herein
that $h_{i}$ are normally distributed with mean $0$ and variance
$\sigma^{2}$. We note that $\sigma/J$ gives us a dimensionless order
parameter to quantify the strength of disorder. 

So far, we have made no restrictions on the decomposition scheme $H=A+B$
other than $\rho^{\left(A\right)}$ and $\rho^{\left(B\right)}$ being
easily computable. A natural question to pose is whether certain choices
of decompositions are intrinsically superior to others. For the Anderson
Hamiltonian, we consider two reasonable partitioning schemes:\begin{subequations}
\begin{equation}
H=A_{1}+B_{1}=\left(\begin{array}{cccc}
h_{1}\\
 & h_{2}\\
 &  & h_{3}\\
 &  &  & \ddots
\end{array}\right)+\left(\begin{array}{cccc}
0 & J\\
J & 0 & J\\
 & J & 0 & \ddots\\
 &  & \ddots & \ddots
\end{array}\right)\label{eq:scheme-i}
\end{equation}
\begin{equation}
H=A_{2}+B_{2}=\left(\begin{array}{ccccc}
h_{1} & J\\
J & 0\\
 &  & h_{3} & J\\
 &  & J & 0\\
 &  &  &  & \ddots
\end{array}\right)+\left(\begin{array}{ccccc}
0\\
 & h_{2} & J\\
 & J & 0\\
 &  &  & h_{4} & \cdots\\
 &  &  & \vdots & \ddots
\end{array}\right).\label{eq:scheme-ii}
\end{equation}
\end{subequations}We refer to these as Scheme~I and II respectively.
For both schemes, each fragment matrix on the right hand side has
a DOS that is easy to determine. In Scheme~I, we have $\rho_{A_{1}}=p_{h}$
since $A_{1}$ is diagonal with each nonzero matrix element being
iid. $B_{1}$ is simply $J$ multiplied by the adjacency matrix of
a one-dimensional chain, and therefore has eigenvalues $\lambda_{n}=2J\cos\left(2n\pi/N\right)$~\cite{Strang1999}.
Then the DOS of $B_{1}$ is $\rho_{B_{1}}\left(\xi\right)=\sum_{n=1}^{N}\delta\left(\xi-\lambda_{n}\right)$
which converges as $N\rightarrow\infty$ to the arcsine distribution
with PDF $p_{AS}\left(\xi\right)=1/\left(\pi\sqrt{4J^{2}-\xi^{2}}\right)$
on the interval $\left[-2\left|J\right|,2\left|J\right|\right]$.
In Scheme~II, we have that $\rho_{A_{2}}=\rho_{B_{2}}=\rho_{X}$
where $\rho_{X}$ is the DOS of $X=\left(\begin{array}{cc}
h_{1} & J\\
J & 0
\end{array}\right)$. Since $X$ has eigenvalues $\epsilon_{\pm}\left(\xi\right)=h_{1}\left(\xi\right)/2\pm\sqrt{h_{1}^{2}\left(\xi\right)/4+J^{2}}$,
their distribution can be calculated to be 
\begin{align}
\rho_{X}\left(\xi\right) & =\left(1+\frac{J^{2}}{\xi^{2}}\right)p_{h}\left(\xi-\frac{J^{2}}{\xi}\right).
\end{align}

\paragraph{Numerical free convolution.---}

\begin{figure}
\caption{\label{fig:dos}Calculation of the DOS, $\rho(\xi)$, of the Hamiltonian
$H$ of (\ref{eq:anderson}) with $M=5000$ samples of $2000\times2000$
matrices for (a) low, (b) moderate and (c) high noise ($\sigma/J$=0.1,
1 and 10 respectively with $\sigma=1$). For each figure we show the
results of free convolution defined in Scheme I ($\rho^{\left(A_{1}\boxplus B_{1}\right)}$;
black solid line), Scheme II ($\rho^{\left(A_{2}\boxplus B_{2}\right)}$;
green dashed line) and exact diagonalization ($\rho^{\left(H\right)}$;
red dotted line).}

\includegraphics[width=8.6cm]{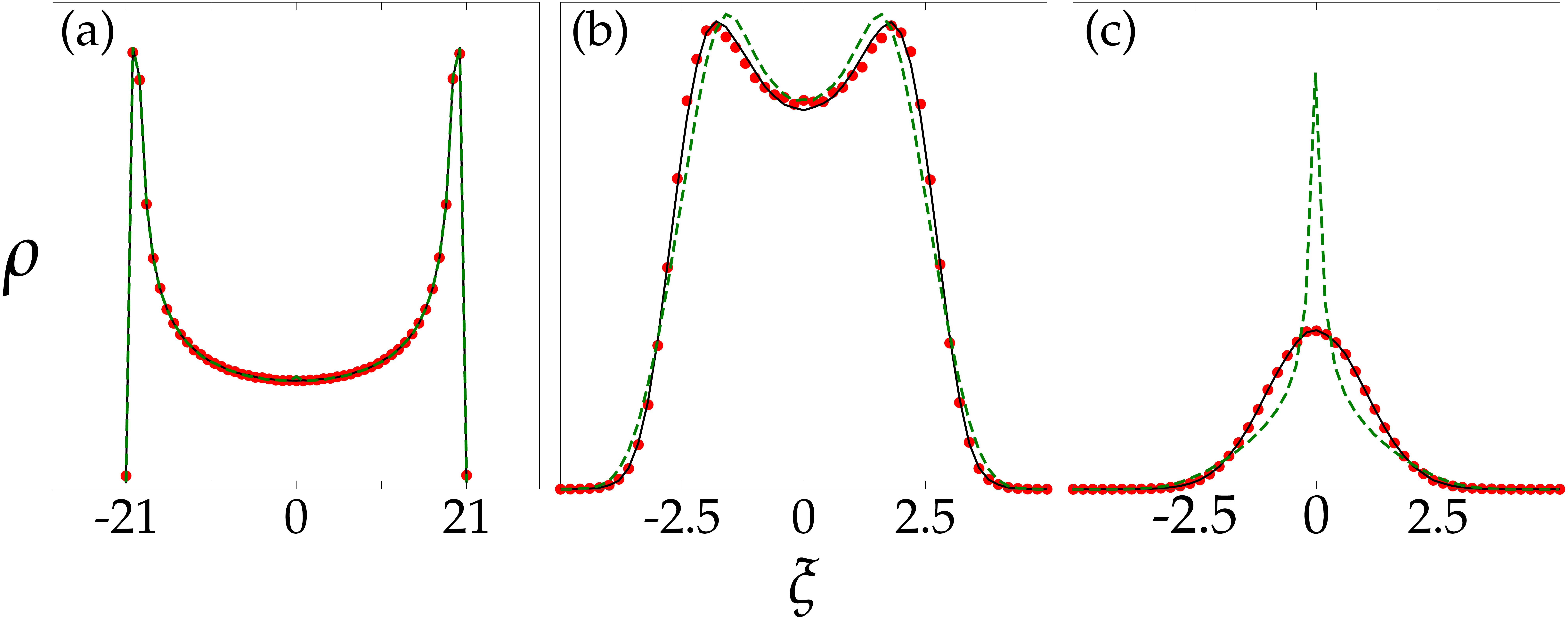}
\end{figure}

We now calculate the free convolution $A\boxplus B$ numerically by
sampling the distributions of $A$ and $B$ and diagonalizing the
free approximant (\ref{eq:free-approximant}). The exact DOS $\rho^{\left(A+B\right)}$
and free approximant $\rho^{\left(A\boxplus B\right)}$ are plotted
in Figure~\ref{fig:dos}(a)--(c) for both schemes for low, moderate
and high noise regimes ($\sigma/J=$0.1, 1, 10 respectively).
We observe that for Scheme~I we have excellent agreement between
$\rho^{\left(H\right)}$ and $\rho^{\left(A_{1}\boxplus B_{1}\right)}$
across all values of $\sigma/J$, which is evident from visual inspection;
in contrast, Scheme~II shows variable quality of fit.

\begin{figure}
\caption{\label{fig:hopping}Diagrammatic expansion of the term $\left\langle A_{1}B_{1}A_{1}B_{1}A_{1}B_{1}A_{1}B_{1}\right\rangle $
in terms of allowed paths dictated by the matrix elements of $A_{1}$
and $B_{1}$ of Scheme~I in~(\ref{eq:scheme-i}).}

\includegraphics[width=8.6cm]{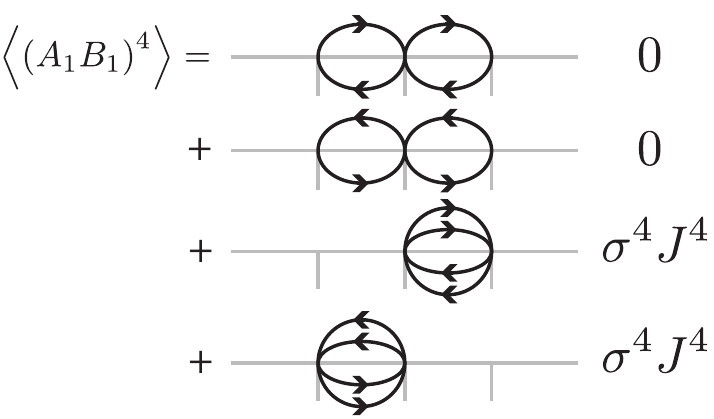}
\end{figure}

We can understand the starkly different behaviors of the two partitioning
schemes using the procedure outlined above to analyze the accuracy
of the approximations $\rho^{\left(H\right)}\approx\rho^{\left(A_{1}\boxplus B_{1}\right)}$
and $\rho^{\left(H\right)}\approx\rho^{\left(A_{2}\boxplus B_{2}\right)}$.
For Scheme~I, we observe that the approximation (\ref{eq:error-term})
is of degree $k=8$; the discrepancy lies solely in the term $\left\langle \left(A_{1}B_{1}\right)^{4}\right\rangle $~\cite{Popescu2011}.
Free probability expects this term to vanish, since both $A_{1}$
and $B_{1}$ are centered (i.e. $\left\langle A_{1}\right\rangle =\left\langle B_{1}\right\rangle =0$)
and hence must satisfy (\ref{eq:free-indep-matrices}) with $n_{1}=m_{1}=\cdots=n_{4}=m_{4}=1$.
In contrast, we can calculate its true value from the definitions
of $A_{1}$ and $B_{1}$. By definition of the NET $\left\langle \cdot\right\rangle $,
only closed paths contribute to the term. Hence, only two types of
terms can contribute to $\left\langle \left(A_{1}B_{1}\right)^{4}\right\rangle $;
these are expressed diagrammatically in Figure~\ref{fig:hopping}.
The matrix $A_{1}$ weights each path by a factor of $h$, while $B_{1}$
weights each path by $J$, and in addition forces the path to hop
to an adjacent site. 
 Consequently, we can write explicitly
\begin{align}
\left\langle \left(A_{1}B_{1}\right)^{4}\right\rangle = & \frac{1}{N}\sum_{i}\mathbb{E}\left(h_{i}Jh_{i-1}Jh_{i}Jh_{i+1}J\right)\nonumber \\
 & +\frac{1}{N}\sum_{i}\mathbb{E}\left(h_{i}Jh_{i+1}Jh_{i}Jh_{i-1}J\right)\nonumber \\
 & +\frac{1}{N}\sum_{i}\mathbb{E}\left(h_{i}Jh_{i-1}Jh_{i}Jh_{i-1}J\right)\nonumber \\
 & +\frac{1}{N}\sum_{i}\mathbb{E}\left(h_{i}Jh_{i+1}Jh_{i}Jh_{i+1}J\right)\nonumber \\
= & 2J^{4}\mathbb{E}\left(h_{i}\right)^{2}\mathbb{E}\left(h_{i}^{2}\right)+2J^{4}\mathbb{E}\left(h_{i}^{2}\right)^{2}=0+2J^{4}\sigma^{4},
\end{align}
where the second equality follows from the independence of the $h_{i}$'s.
As this is the only source of discrepancy at the eighth moment, this
explains why the agreement between the free and exact PDFs is so good,
as the leading order correction is in the eighth derivative of $\rho^{\left(A_{1}\boxplus B_{1}\right)}$
with coefficient $2\sigma^{4}J^{4}/8!=\left(\sigma J\right)^{4}/20160$.
In contrast, we observe for Scheme~II that the leading order correction
is at $k=4$, where the discrepancy lies in $\left\langle A_{2}^{2}B_{2}^{2}\right\rangle $.
Free probability expects this to be equal to $\left\langle A_{2}^{2}B_{2}^{2}\right\rangle =\left\langle A_{2}^{2}\right\rangle \left\langle B_{2}^{2}\right\rangle =\left\langle X^{2}\right\rangle ^{2}=\left(J^{2}+\sigma^{2}/2\right)^{2}$,
whereas the exact value of this term is $J^{2}\left(J^{2}+\sigma^{2}\right)$.
Therefore the discrepancy is in the fourth derivative of $\rho^{\left(A\boxplus B\right)}$
with coefficient $\left(-\sigma^{4}/4\right)/4!=-\sigma^{4}/96$.

\paragraph{Analytic free convolution.---}

Free probability allows us also to calculate the limiting distributions
of $\rho^{\left(A\boxplus B\right)}$ in the macroscopic limit of
infinite matrix sizes $N\rightarrow\infty$ and infinite samples $M\rightarrow\infty$.
In this limit, the DOS $\rho^{\left(A\boxplus B\right)}$ is given
as a particular type of integral convolution of $\rho^{\left(A\right)}$
and $\rho^{\left(B\right)}$. We now calculate the free convolution
analytically in the macroscopic limit for the two partitioning schemes
discussed above, thus sidestepping the cost of sampling and matrix
diagonalization altogether.

The key tool to performing the free convolution analytically is the
$R$-transform $r\left(w\right)=g^{-1}\left(w\right)-w^{-1}$~\cite{Voiculescu1985},
where $g^{-1}$ is defined implicitly via the Cauchy transform
\begin{equation}
w=\int_{\mathbb{R}}\frac{\rho\left(\xi\right)}{g^{-1}\left(w\right)-\xi}d\xi.
\end{equation}
For freely independent $A$ and $B$, the $R$-transforms linearize
the free convolution, i.e. $R^{\left(A\boxplus B\right)}\left(w\right)=R^{\left(A\right)}\left(w\right)+R^{\left(B\right)}\left(w\right)$,
and that the PDF can be recovered from the Plemelj--Sokhotsky inversion
formula by\begin{subequations} 
\begin{align}
\rho^{\left(A\boxplus B\right)}\left(\xi\right) & =\frac{1}{\pi}\mbox{Im}\left(\left(g^{\left(A\boxplus B\right)}\right)^{-1}\left(\xi\right)\right)\\
g^{\left(A\boxplus B\right)}\left(w\right) & =R^{\left(A\boxplus B\right)}\left(w\right)+w^{-1}.
\end{align}
\end{subequations}As an example, we apply this to Scheme~I with
each iid $h_{i}$ following a Wigner semicircle distribution with
PDF $p_{W}\left(\xi\right)=\sqrt{4-\xi^{2}}/4\pi$ on the interval
$\left[-2,2\right]$. As described earlier (Using semicircular noise
instead of Gaussian noise simplifies the analytic calculation considerably.)
From the DOS $\rho^{\left(A_{1}\right)}=p_{W}$, we calculate its
Cauchy transform (i.e. its retarded Green function)
\begin{align}
G^{\left(A_{1}\right)}\left(z\right) & =\lim_{\epsilon\downarrow0}\int_{\mathbb{R}}\frac{\rho^{\left(A_{1}\right)}\left(\xi\right)}{z-\left(\xi+i\epsilon\right)}d\xi=\frac{z-\sqrt{z^{2}-4}}{2}.
\end{align}
Next, take the functional inverse 
\begin{equation}
g^{\left(A_{1}\right)}\left(w\right)=\left(G^{\left(A_{1}\right)}\right)^{-1}\left(w\right)=w+\frac{1}{w}.
\end{equation}
Subtracting $1/w$ finally yields the $R$-transform $ $$r^{\left(A\right)}\left(w\right)=w$.
Similarly with $\rho^{\left(B_{1}\right)}=p_{AS}$, we have its Cauchy
transform
\begin{equation}
G^{\left(B_{1}\right)}\left(z\right)=\lim_{\epsilon\downarrow0}\int_{\mathbb{R}}\frac{\rho^{\left(B_{1}\right)}\left(\xi\right)}{z-\left(\xi+i\epsilon\right)}d\xi=\frac{1}{\sqrt{z^{2}-4J^{2}}}
\end{equation}
and its functional inverse
\begin{equation}
g^{\left(B_{1}\right)}\left(w\right)=\frac{\sqrt{1+4J^{2}w^{2}}}{w},
\end{equation}
which finally yields the $R$-transform $R^{\left(B_{1}\right)}\left(w\right)=\left(-1+\sqrt{1+4J^{2}w^{2}}\right)/w$.

\begin{figure}
\caption{\label{fig:dos-semicircle}DOS, $\rho(\xi)$, of the Hamiltonian~(\ref{eq:anderson})
with $M=5000$ samples of $2000\times2000$ matrices with (a) low,
(b) moderate and (c) high semicircular on-site noise ($\sigma/J$=0.1,
1 and 10 respectively with $\sigma=1$), as calculated with exact
diagonalization (red dotted line), free convolution (black solid line),
and perturbation theory with $A_{1}$ as reference (blue dashed line)
and $B_{1}$ as reference (gray dash-dotted line). The partitioning
scheme is Scheme~I of~(\ref{eq:scheme-i}).}

\centering{}\includegraphics[width=8.6cm]{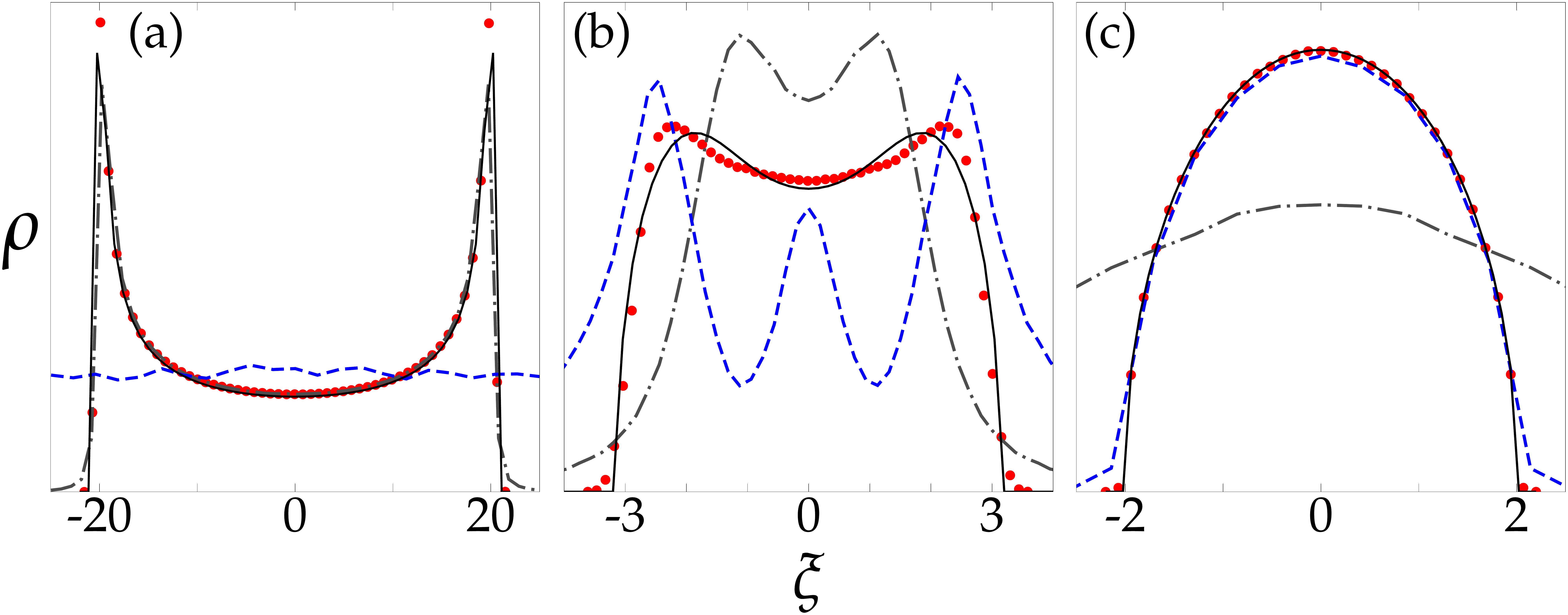}
\end{figure}

To perform the free convolution analytically, we add the $R$-transforms
to get $R^{\left(A_{1}\boxplus B_{1}\right)}\left(w\right)=R^{\left(A_{1}\right)}\left(w\right)+R^{\left(B_{1}\right)}\left(w\right)$,
from which we obtain
\begin{equation}
g^{\left(A_{1}\boxplus B_{1}\right)}\left(w\right)=w+\frac{\sqrt{1+4J^{2}w^{2}}}{w}.
\end{equation}
The final steps are to calculate the functional inverse $\left(g^{\left(A_{1}\boxplus B_{1}\right)}\right)^{-1}$
and take its imaginary part to obtain $\rho^{\left(A_{1}\boxplus B_{1}\right)}$.
Unfortunately, $\left(g^{\left(A_{1}\boxplus B_{1}\right)}\right)^{-1}$
cannot be written in a compact closed form; nevertheless, the inversion
can be calculated numerically. We present calculations of the DOS
as a function of noise strength $\sigma/J$ in Figure~\ref{fig:dos-semicircle},
showing again that the free convolution is an excellent approximation
to the exact DOS.

\paragraph{Comparison with other approximations.---}

For comparative purposes, we also performed calculations using standard
second-order matrix perturbation theory~\cite{Horn1990} for both
partitioning schemes. The results are also shown in Figure~\ref{fig:dos-semicircle}.
Unsurprisingly, perturbation theory produces results that vary strongly
with $\sigma/J$, and that the different series, based on whether
$A$ is considered a perturbation of $B$ or vice versa, have different
regimes of applicability. Furthermore it is clear even from visual
inspection that the second moment of the DOS calculated using second-order
perturbation theory is in general incorrect. In contrast, the free
convolution produces results with a more uniform level of accuracy
across the entire range of $\sigma/J$, and that we have at least
the first three moments being correct~\cite{Movassagh2010}.

It is also natural to ask what mean-field theory, another standard
tool, would predict. Interestingly, the limiting behavior of Scheme~I
as $N\rightarrow\infty$ is equivalent to a form of mean-field theory
known as the coherent potential approximation (CPA)~\cite{Neu1994,Neu1995a,Neu1995b}
in condensed matter physics, and is equivalent to the Blue's function
formalism in quantum chromodynamics for calculating one-particle irreducible
self-energies~\cite{Zee1996a}. The breakdown in the CPA in the term
$\left\langle \left(A_{1}B_{1}\right)^{4}\right\rangle $ is known~\cite{Blackman1971,Thouless1974};
however, to our knowledge, the magnitude of the deviation was not
explained. In contrast, our error analysis framework affords us such
a quantitative explanation.

Finally, we discuss the predictions of isotropic entanglement theory,
which proposes a linear interpolation between the classical convolution
$\rho^{\left(A*B\right)}\left(\xi\right)=\int_{-\infty}^{\infty}\rho^{\left(A\right)}\left(\xi\right)\rho^{\left(B\right)}\left(x-\xi\right)dx$
and the free convolution $\rho^{\left(A\boxplus B\right)}\left(\xi\right)$
in the fourth cumulant~\cite{Movassagh2010,Movassagh2011a}. The
classical convolution can be calculated directly from the random matrices
$A$ and $B$; by diagonalizing the matrices as $A=Q_{A}^{-1}\Lambda_{A}Q_{A}$
and $B=Q_{B}^{-1}\Lambda_{B}Q_{B}$, the classical convolution $\rho^{\left(A*B\right)}\left(\xi\right)$
can be computed from the eigenvalues of random matrices of the form
$Z_{cl}=\Lambda_{A}+\Pi^{-1}\Lambda_{B}\Pi$ where $\Pi$ is a $N\times N$
random permutation matrix. It is instructive to compare this with
the free convolution, which can be sampled from matrices of the form
$Z^{\prime}=\Lambda_{A}+Q^{-1}\Lambda_{B}Q$, which can be shown by
orthogonal invariance of the Haar measure random matrices $Q$ to
be equivalent to sampling matrices of the form $Z=A+Q^{-1}BQ$ described
previously.

As discussed previously, the lowest three moments of $Z$ and $H$
are identical; this turns out to be true also for $Z_{cl}$~\cite{Movassagh2010}.
Therefore IE proposes to interpolate via the fourth cumulant, with
interpolation parameter $p$ defined as
\begin{equation}
p=\frac{\kappa_{4}^{\left(H\right)}-\kappa_{4}^{\left(A\boxplus B\right)}}{\kappa_{4}^{\left(A*B\right)}-\kappa_{4}^{\left(A\boxplus B\right)}}
\end{equation}

We observe that for Scheme~I, IE appears to always favor the free
convolution limit ($p=0$) as opposed to the classical limit ($p=1$);
this is not surprising as we know from our previous analysis that
$\kappa_{4}^{\left(H\right)}=\kappa_{4}^{\left(A_{1}\boxplus B_{1}\right)}$,
and that the agreement with the exact diagonalization result is excellent
regardless of $\sigma/J$. In Scheme~II, however, we observe the
unexpected result that $p$ is always negative and that the agreement
varies with the noise strength $\sigma/J$. From the moment expansion
we understand why; we have that the first three moments match while
$\kappa_{4}^{\left(A_{2}+B_{2}\right)}-\kappa_{4}^{\left(A_{2}\boxplus B_{2}\right)}=-\sigma^{4}/4$.
The discrepancy lies in the term $\left\langle A_{2}^{2}B_{2}^{2}\right\rangle $,
which is expected to have the value $\left\langle A_{2}^{2}\right\rangle \left\langle B_{2}^{2}\right\rangle =\left(J^{2}+\sigma^{2}/2\right)^{2}$
in free probability but instead has the exact value $J^{2}\left(J^{2}+\sigma^{2}\right)$.
Furthermore, we have that $\kappa_{4}^{\left(A_{2}*B_{2}\right)}\ne\kappa_{4}^{\left(A_{2}\boxplus B_{2}\right)}$where
the only discrepancy lies is in the so-called departing term $\left\langle A_{2}B_{2}A_{2}B_{2}\right\rangle $~\cite{Movassagh2010,Movassagh2011a}.
This term contributes 0 to $\kappa_{4}^{\left(A\boxplus B\right)}$
but has value $\left\langle A_{2}^{2}\right\rangle \left\langle B_{2}^{2}\right\rangle =\left(J^{2}+\sigma^{2}/2\right)^{2}$
in $\kappa_{4}^{\left(A_{2}*B_{2}\right)}$, since for the classical
convolution we have that $\left\langle \Pi_{s=1}^{r}\left(A_{2}^{n_{s}}B_{2}^{m_{s}}\right)\right\rangle =\left\langle A_{2}^{\sum_{s=1}^{r}n_{s}}\right\rangle \left\langle B_{2}^{\sum_{s=1}^{r}m_{s}}\right\rangle $.
This therefore explains why we observe a negative $p$, as this calculation
shows that
\begin{equation}
p=\frac{\kappa_{4}^{\left(A_{2}+B_{2}\right)}-\kappa_{4}^{\left(A_{2}\boxplus B_{2}\right)}}{\kappa_{4}^{\left(A_{2}*B_{2}\right)}-\kappa_{4}^{\left(A_{2}\boxplus B_{2}\right)}}=-2\left(2\left(\frac{\sigma}{J}\right)^{-2}+1\right)^{-2}
\end{equation}
which is manifestly negative.

In conclusion, we have demonstrated that the accuracy of approximations
using the free convolution depend crucially on the particular choice
of partitioning scheme for the Hamiltonian. We have found an unexpectedly
accurate approximation for the DOS of disordered Hamiltonians, both
for finite dimensional systems and in the macroscopic limit $N\rightarrow\infty$.
In particular, this approximation remains accurate no matter the strength
of noise present in the system. Our error analysis framework provides
an explanation for this accuracy, namely that the lowest seven moments
of the eigenvalues distribution are correct, with the first discrepancy
only in one particular term arising at the eighth moment. 

We expect our results to be generally applicable to arbitrary Hamiltonians,
and are currently investigating the validity of these approximations
for electronic structure models on two- and three-dimensional lattices.
These results pave the way toward constructing even more accurate
approximations using free probability, guided by a rigorous error
analysis framework in terms of the accuracy of successive moments.
Our results represent an optimistic beginning to the use of powerful
and highly accurate nonperturbative methods for studying the electronic
properties of disordered condensed matter systems regardless of the
strength of noise present. We expect that these methods will be especially
useful when the presence of noise is not merely a perturbation of
a perfect system, but rather, crucial to the emergence of unique physical
phenomena.
\begin{acknowledgments}
J.C., E.H., M.W., T.V., R.M., and A.E. acknowledge funding from NSF
SOLAR Grant No.~1035400. J.M.~acknowledges support from NSF CHE
Grant No. 1112825 and DARPA Grant No. N99001-10-1-4063. A.E.~acknowledges
additional funding from NSF DMS Grant No.~1016125. A.S.~acknowledges
funding from Spain's Direcci\'on General de Investigaci\'on, Project
TIN2010-21575-C02-02. We thank Jonathan Novak (MIT), Sebastiaan Vlaming (MIT)
and Raj Rao (Michigan) for insightful discussions.
\end{acknowledgments}
\bibliographystyle{apsrev4-1}
\bibliography{cpa,general,localization,moment_expansion,rmt,intro}

\begin{thebibliography}{34}%
\makeatletter
\providecommand \@ifxundefined [1]{%
 \@ifx{#1\undefined}
}%
\providecommand \@ifnum [1]{%
 \ifnum #1\expandafter \@firstoftwo
 \else \expandafter \@secondoftwo
 \fi
}%
\providecommand \@ifx [1]{%
 \ifx #1\expandafter \@firstoftwo
 \else \expandafter \@secondoftwo
 \fi
}%
\providecommand \natexlab [1]{#1}%
\providecommand \enquote  [1]{``#1''}%
\providecommand \bibnamefont  [1]{#1}%
\providecommand \bibfnamefont [1]{#1}%
\providecommand \citenamefont [1]{#1}%
\providecommand \href@noop [0]{\@secondoftwo}%
\providecommand \href [0]{\begingroup \@sanitize@url \@href}%
\providecommand \@href[1]{\@@startlink{#1}\@@href}%
\providecommand \@@href[1]{\endgroup#1\@@endlink}%
\providecommand \@sanitize@url [0]{\catcode `\\12\catcode `\$12\catcode
  `\&12\catcode `\#12\catcode `\^12\catcode `\_12\catcode `\%12\relax}%
\providecommand \@@startlink[1]{}%
\providecommand \@@endlink[0]{}%
\providecommand \url  [0]{\begingroup\@sanitize@url \@url }%
\providecommand \@url [1]{\endgroup\@href {#1}{\urlprefix }}%
\providecommand \urlprefix  [0]{URL }%
\providecommand \Eprint [0]{\href }%
\providecommand \doibase [0]{http://dx.doi.org/}%
\providecommand \selectlanguage [0]{\@gobble}%
\providecommand \bibinfo  [0]{\@secondoftwo}%
\providecommand \bibfield  [0]{\@secondoftwo}%
\providecommand \translation [1]{[#1]}%
\providecommand \BibitemOpen [0]{}%
\providecommand \bibitemStop [0]{}%
\providecommand \bibitemNoStop [0]{.\EOS\space}%
\providecommand \EOS [0]{\spacefactor3000\relax}%
\providecommand \BibitemShut  [1]{\csname bibitem#1\endcsname}%
\let\auto@bib@innerbib\@empty
\bibitem [{\citenamefont {Thouless}(1974)}]{Thouless1974}%
  \BibitemOpen
  \bibfield  {author} {\bibinfo {author} {\bibfnamefont {D.~J.}\ \bibnamefont
  {Thouless}},\ }\href {\doibase 10.1016/0370-1573(74)90029-5} {\bibfield
  {journal} {\bibinfo  {journal} {Phys. Rep.}\ }\textbf {\bibinfo {volume}
  {13}},\ \bibinfo {pages} {93} (\bibinfo {year} {1974})}\BibitemShut {NoStop}%
\bibitem [{\citenamefont {Evers}\ and\ \citenamefont
  {Mirlin}(2008)}]{Evers:2008p706}%
  \BibitemOpen
  \bibfield  {author} {\bibinfo {author} {\bibfnamefont {F.}~\bibnamefont
  {Evers}}\ and\ \bibinfo {author} {\bibfnamefont {A.}~\bibnamefont {Mirlin}},\
  }\href {\doibase 10.1103/RevModPhys.80.1355} {\bibfield  {journal} {\bibinfo
  {journal} {Rev. Mod. Phys}\ }\textbf {\bibinfo {volume} {80}},\ \bibinfo
  {pages} {1355} (\bibinfo {year} {2008})}\BibitemShut {NoStop}%
\bibitem [{\citenamefont {Bouchaud}\ and\ \citenamefont
  {Georges}(1990)}]{Bouchaud1990}%
  \BibitemOpen
  \bibfield  {author} {\bibinfo {author} {\bibfnamefont {J.-P.}\ \bibnamefont
  {Bouchaud}}\ and\ \bibinfo {author} {\bibfnamefont {A.}~\bibnamefont
  {Georges}},\ }\href {\doibase 10.1016/0370-1573(90)90099-N} {\bibfield
  {journal} {\bibinfo  {journal} {Phys. Rep.}\ }\textbf {\bibinfo {volume}
  {195}},\ \bibinfo {pages} {127} (\bibinfo {year} {1990})}\BibitemShut
  {NoStop}%
\bibitem [{\citenamefont {Shlesinger}\ \emph {et~al.}(1993)\citenamefont
  {Shlesinger}, \citenamefont {Zaslavsky},\ and\ \citenamefont
  {Klafter}}]{Shlesinger1993}%
  \BibitemOpen
  \bibfield  {author} {\bibinfo {author} {\bibfnamefont {M.~F.}\ \bibnamefont
  {Shlesinger}}, \bibinfo {author} {\bibfnamefont {G.~M.}\ \bibnamefont
  {Zaslavsky}}, \ and\ \bibinfo {author} {\bibfnamefont {J.}~\bibnamefont
  {Klafter}},\ }\href {\doibase 10.1038/363031a0} {\bibfield  {journal}
  {\bibinfo  {journal} {Nature}\ }\textbf {\bibinfo {volume} {363}},\ \bibinfo
  {pages} {31} (\bibinfo {year} {1993})}\BibitemShut {NoStop}%
\bibitem [{\citenamefont {Palmer}(1982)}]{Palmer1982}%
  \BibitemOpen
  \bibfield  {author} {\bibinfo {author} {\bibfnamefont {R.~G.}\ \bibnamefont
  {Palmer}},\ }\href {\doibase 10.1080/00018738200101438} {\bibfield  {journal}
  {\bibinfo  {journal} {Adv. Phys.}\ }\textbf {\bibinfo {volume} {31}},\
  \bibinfo {pages} {669} (\bibinfo {year} {1982})}\BibitemShut {NoStop}%
\bibitem [{\citenamefont {Barkai}\ \emph {et~al.}(2004)\citenamefont {Barkai},
  \citenamefont {Jung},\ and\ \citenamefont {Silbey}}]{Barkai2004}%
  \BibitemOpen
  \bibfield  {author} {\bibinfo {author} {\bibfnamefont {E.}~\bibnamefont
  {Barkai}}, \bibinfo {author} {\bibfnamefont {Y.}~\bibnamefont {Jung}}, \ and\
  \bibinfo {author} {\bibfnamefont {R.}~\bibnamefont {Silbey}},\ }\href
  {\doibase 10.1146/annurev.physchem.55.111803.143246} {\bibfield  {journal}
  {\bibinfo  {journal} {Annu. Rev. Phys. Chem.}\ }\textbf {\bibinfo {volume}
  {55}},\ \bibinfo {pages} {457} (\bibinfo {year} {2004})}\BibitemShut
  {NoStop}%
\bibitem [{\citenamefont {Stefani}\ \emph {et~al.}(2009)\citenamefont
  {Stefani}, \citenamefont {Hoogenboom},\ and\ \citenamefont
  {Barkai}}]{Stefani2009}%
  \BibitemOpen
  \bibfield  {author} {\bibinfo {author} {\bibfnamefont {F.~D.}\ \bibnamefont
  {Stefani}}, \bibinfo {author} {\bibfnamefont {J.~P.}\ \bibnamefont
  {Hoogenboom}}, \ and\ \bibinfo {author} {\bibfnamefont {E.}~\bibnamefont
  {Barkai}},\ }\href {\doibase 10.1063/1.3086100} {\bibfield  {journal}
  {\bibinfo  {journal} {Phys. Today}\ }\textbf {\bibinfo {volume} {62}},\
  \bibinfo {pages} {34} (\bibinfo {year} {2009})}\BibitemShut {NoStop}%
\bibitem [{\citenamefont {Hernando}(1999)}]{Hernando1999}%
  \BibitemOpen
  \bibfield  {author} {\bibinfo {author} {\bibfnamefont {A.}~\bibnamefont
  {Hernando}},\ }\href {\doibase 10.1088/0953-8984/11/48/308} {\bibfield
  {journal} {\bibinfo  {journal} {J. Phys.: Condens. Matter}\ }\textbf
  {\bibinfo {volume} {11}},\ \bibinfo {pages} {9455} (\bibinfo {year}
  {1999})}\BibitemShut {NoStop}%
\bibitem [{\citenamefont {Dyre}\ and\ \citenamefont
  {Schr{\o}der}(2000)}]{Dyre2000}%
  \BibitemOpen
  \bibfield  {author} {\bibinfo {author} {\bibfnamefont {J.}~\bibnamefont
  {Dyre}}\ and\ \bibinfo {author} {\bibfnamefont {T.}~\bibnamefont
  {Schr{\o}der}},\ }\href {\doibase 10.1103/RevModPhys.72.873} {\bibfield
  {journal} {\bibinfo  {journal} {Rev. Mod. Phys.}\ }\textbf {\bibinfo {volume}
  {72}},\ \bibinfo {pages} {873} (\bibinfo {year} {2000})}\BibitemShut
  {NoStop}%
\bibitem [{\citenamefont {Dugdale}(2005)}]{Dugdale2005}%
  \BibitemOpen
  \bibfield  {author} {\bibinfo {author} {\bibfnamefont {J.~S.}\ \bibnamefont
  {Dugdale}},\ }\href {\doibase 10.2277/0521017513} {\emph {\bibinfo {title}
  {{The Electrical Properties of Disordered Metals}}}},\ Cambridge Solid State
  Science Series\ (\bibinfo  {publisher} {Cambridge},\ \bibinfo {address}
  {Cambridge, UK},\ \bibinfo {year} {2005})\BibitemShut {NoStop}%
\bibitem [{\citenamefont {Peet}\ \emph {et~al.}(2009)\citenamefont {Peet},
  \citenamefont {Heeger},\ and\ \citenamefont {Bazan}}]{Peet2009}%
  \BibitemOpen
  \bibfield  {author} {\bibinfo {author} {\bibfnamefont {J.}~\bibnamefont
  {Peet}}, \bibinfo {author} {\bibfnamefont {A.~J.}\ \bibnamefont {Heeger}}, \
  and\ \bibinfo {author} {\bibfnamefont {G.~C.}\ \bibnamefont {Bazan}},\ }\href
  {\doibase 10.1021/ar900065j} {\bibfield  {journal} {\bibinfo  {journal} {Acc.
  Chem. Res.}\ }\textbf {\bibinfo {volume} {42}},\ \bibinfo {pages} {1700}
  (\bibinfo {year} {2009})}\BibitemShut {NoStop}%
\bibitem [{\citenamefont {Difley}\ \emph {et~al.}(2010)\citenamefont {Difley},
  \citenamefont {Wang}, \citenamefont {Yeganeh}, \citenamefont {Yost},\ and\
  \citenamefont {{Van Voorhis}}}]{Difley2010}%
  \BibitemOpen
  \bibfield  {author} {\bibinfo {author} {\bibfnamefont {S.}~\bibnamefont
  {Difley}}, \bibinfo {author} {\bibfnamefont {L.-P.}\ \bibnamefont {Wang}},
  \bibinfo {author} {\bibfnamefont {S.}~\bibnamefont {Yeganeh}}, \bibinfo
  {author} {\bibfnamefont {S.~R.}\ \bibnamefont {Yost}}, \ and\ \bibinfo
  {author} {\bibfnamefont {T.}~\bibnamefont {{Van Voorhis}}},\ }\href@noop {}
  {\bibfield  {journal} {\bibinfo  {journal} {Acc. Chem. Res.}\ }\textbf
  {\bibinfo {volume} {43}},\ \bibinfo {pages} {995} (\bibinfo {year}
  {2010})}\BibitemShut {NoStop}%
\bibitem [{\citenamefont {Yost}\ \emph {et~al.}(2011)\citenamefont {Yost},
  \citenamefont {Wang},\ and\ \citenamefont {{Van Voorhis}}}]{Yost2011}%
  \BibitemOpen
  \bibfield  {author} {\bibinfo {author} {\bibfnamefont {S.~R.}\ \bibnamefont
  {Yost}}, \bibinfo {author} {\bibfnamefont {L.-P.}\ \bibnamefont {Wang}}, \
  and\ \bibinfo {author} {\bibfnamefont {T.}~\bibnamefont {{Van Voorhis}}},\
  }\href {\doibase 10.1021/jp203387m} {\bibfield  {journal} {\bibinfo
  {journal} {J. Phys. Chem. C}\ }\textbf {\bibinfo {volume} {115}},\ \bibinfo
  {pages} {14431} (\bibinfo {year} {2011})}\BibitemShut {NoStop}%
\bibitem [{\citenamefont {Kollman}(1993)}]{Kollman1993}%
  \BibitemOpen
  \bibfield  {author} {\bibinfo {author} {\bibfnamefont {P.}~\bibnamefont
  {Kollman}},\ }\href {\doibase 10.1021/cr00023a004} {\bibfield  {journal}
  {\bibinfo  {journal} {Chem. Rev.}\ }\textbf {\bibinfo {volume} {93}},\
  \bibinfo {pages} {2395} (\bibinfo {year} {1993})}\BibitemShut {NoStop}%
\bibitem [{\citenamefont {Voiculescu}(1991)}]{Voiculescu1991}%
  \BibitemOpen
  \bibfield  {author} {\bibinfo {author} {\bibfnamefont {D.}~\bibnamefont
  {Voiculescu}},\ }\href {\doibase 10.1007/BF01245072} {\bibfield  {journal}
  {\bibinfo  {journal} {Invent. Math.}\ }\textbf {\bibinfo {volume} {104}},\
  \bibinfo {pages} {201} (\bibinfo {year} {1991})}\BibitemShut {NoStop}%
\bibitem [{\citenamefont {Biane}(1998)}]{Biane1998}%
  \BibitemOpen
  \bibfield  {author} {\bibinfo {author} {\bibfnamefont {P.}~\bibnamefont
  {Biane}},\ }in\ \href
  {http://citeseerx.ist.psu.edu/viewdoc/download?doi=10.1.1.43.6806\&amp;rep=rep1\&amp;type=pdf}
  {\emph {\bibinfo {booktitle} {Quantum probability communications}}},\
  Vol.~\bibinfo {volume} {11}\ (\bibinfo {year} {1998})\ Chap.~\bibinfo
  {chapter} {3}, pp.\ \bibinfo {pages} {55--71}\BibitemShut {NoStop}%
\bibitem [{\citenamefont {Zee}(1996)}]{Zee1996a}%
  \BibitemOpen
  \bibfield  {author} {\bibinfo {author} {\bibfnamefont {A.}~\bibnamefont
  {Zee}},\ }\href {\doibase 10.1016/0550-3213(96)00276-3} {\bibfield  {journal}
  {\bibinfo  {journal} {Nucl. Phys. B}\ }\textbf {\bibinfo {volume} {474}},\
  \bibinfo {pages} {726} (\bibinfo {year} {1996})}\BibitemShut {NoStop}%
\bibitem [{\citenamefont {Stuart}\ and\ \citenamefont
  {Ord}(1994)}]{Stuart1994}%
  \BibitemOpen
  \bibfield  {author} {\bibinfo {author} {\bibfnamefont {A.}~\bibnamefont
  {Stuart}}\ and\ \bibinfo {author} {\bibfnamefont {J.~K.}\ \bibnamefont
  {Ord}},\ }\href@noop {} {\emph {\bibinfo {title} {{Kendall's advanced theory
  of statistics.}}}}\ (\bibinfo  {publisher} {Edward Arnold},\ \bibinfo
  {address} {London},\ \bibinfo {year} {1994})\BibitemShut {NoStop}%
\bibitem [{\citenamefont {Blinnikov}\ and\ \citenamefont
  {Moessner}(1998)}]{Blinnikov1998a}%
  \BibitemOpen
  \bibfield  {author} {\bibinfo {author} {\bibfnamefont {S.}~\bibnamefont
  {Blinnikov}}\ and\ \bibinfo {author} {\bibfnamefont {R.}~\bibnamefont
  {Moessner}},\ }\href {\doibase 10.1051/aas:1998221} {\bibfield  {journal}
  {\bibinfo  {journal} {Astron. Astrophys. Supp. Ser.}\ }\textbf {\bibinfo
  {volume} {130}},\ \bibinfo {pages} {193} (\bibinfo {year}
  {1998})}\BibitemShut {NoStop}%
\bibitem [{\citenamefont {Wallace}(1958)}]{Wallace1958}%
  \BibitemOpen
  \bibfield  {author} {\bibinfo {author} {\bibfnamefont {D.}~\bibnamefont
  {Wallace}},\ }\href {http://www.jstor.org/stable/10.2307/2237255} {\bibfield
  {journal} {\bibinfo  {journal} {Ann. Math. Stat.}\ }\textbf {\bibinfo
  {volume} {29}},\ \bibinfo {pages} {635} (\bibinfo {year} {1958})}\BibitemShut
  {NoStop}%
\bibitem [{\citenamefont {Nica}\ and\ \citenamefont
  {Speicher}(2006)}]{Nica2006a}%
  \BibitemOpen
  \bibfield  {author} {\bibinfo {author} {\bibfnamefont {A.}~\bibnamefont
  {Nica}}\ and\ \bibinfo {author} {\bibfnamefont {R.}~\bibnamefont
  {Speicher}},\ }\href@noop {} {\emph {\bibinfo {title} {{Lectures on the
  Combinatorics of Free Probability}}}},\ London Math. Soc. Lecture Note Ser.\
  (\bibinfo {address} {London},\ \bibinfo {year} {2006})\BibitemShut {NoStop}%
\bibitem [{\citenamefont {Diaconis}(2005)}]{Diaconis2005}%
  \BibitemOpen
  \bibfield  {author} {\bibinfo {author} {\bibfnamefont {P.}~\bibnamefont
  {Diaconis}},\ }\href@noop {} {\bibfield  {journal} {\bibinfo  {journal} {Not.
  Amer. Math. Soc.}\ }\textbf {\bibinfo {volume} {52}},\ \bibinfo {pages}
  {1348} (\bibinfo {year} {2005})}\BibitemShut {NoStop}%
\bibitem [{\citenamefont {Voiculescu}(1994)}]{Voiculescu1994}%
  \BibitemOpen
  \bibfield  {author} {\bibinfo {author} {\bibfnamefont {D.}~\bibnamefont
  {Voiculescu}},\ }in\ \href
  {http://www.mathunion.org/ICM/ICM1994.1/Main/icm1994.1.0227.0242.ocr.pdf}
  {\emph {\bibinfo {booktitle} {Proceedings of the International Congress of
  Mathematicians}}}\ (\bibinfo  {publisher} {Birkh{\"a}user Verlag},\ \bibinfo
  {address} {Z{\"u}rich, Switzerland},\ \bibinfo {year} {1994})\ pp.\ \bibinfo
  {pages} {227--241}\BibitemShut {NoStop}%
\bibitem [{\citenamefont {Sawada}(2001)}]{Sawada2001}%
  \BibitemOpen
  \bibfield  {author} {\bibinfo {author} {\bibfnamefont {J.}~\bibnamefont
  {Sawada}},\ }\href {\doibase 10.1137/S0097539700377037} {\bibfield  {journal}
  {\bibinfo  {journal} {SIAM J. Comput.}\ }\textbf {\bibinfo {volume} {31}},\
  \bibinfo {pages} {259} (\bibinfo {year} {2001})}\BibitemShut {NoStop}%
\bibitem [{\citenamefont {Movassagh}\ and\ \citenamefont
  {Edelman}()}]{Movassagh2010}%
  \BibitemOpen
  \bibfield  {author} {\bibinfo {author} {\bibfnamefont {R.}~\bibnamefont
  {Movassagh}}\ and\ \bibinfo {author} {\bibfnamefont {A.}~\bibnamefont
  {Edelman}},\ }\href {http://arxiv.org/abs/1012.5039} {\ }\Eprint
  {http://arxiv.org/abs/1012.5039} {arXiv:1012.5039} \BibitemShut {NoStop}%
\bibitem [{\citenamefont {Strang}(1999)}]{Strang1999}%
  \BibitemOpen
  \bibfield  {author} {\bibinfo {author} {\bibfnamefont {G.}~\bibnamefont
  {Strang}},\ }\href {\doibase 10.1137/S0036144598336745} {\bibfield  {journal}
  {\bibinfo  {journal} {SIAM Rev.}\ }\textbf {\bibinfo {volume} {41}},\
  \bibinfo {pages} {135} (\bibinfo {year} {1999})}\BibitemShut {NoStop}%
\bibitem [{\citenamefont {Popescu}(2011)}]{Popescu2011}%
  \BibitemOpen
  \bibfield  {author} {\bibinfo {author} {\bibfnamefont {I.}~\bibnamefont
  {Popescu}},\ }\href@noop {} {}\bibinfo {howpublished} {personal
  communication} (\bibinfo {year} {2011})\BibitemShut {NoStop}%
\bibitem [{\citenamefont {Voiculescu}(1985)}]{Voiculescu1985}%
  \BibitemOpen
  \bibfield  {author} {\bibinfo {author} {\bibfnamefont {D.}~\bibnamefont
  {Voiculescu}},\ }in\ \href {\doibase 10.1007/BFb0074909} {\emph {\bibinfo
  {booktitle} {Operator algebras and their connections with topology and
  ergodic theory}}},\ \bibinfo {series} {Lecture Notes in Mathematics}, Vol.\
  \bibinfo {volume} {1132},\ \bibinfo {editor} {edited by\ \bibinfo {editor}
  {\bibfnamefont {H.}~\bibnamefont {Araki}}, \bibinfo {editor} {\bibfnamefont
  {C.}~\bibnamefont {Moore}}, \bibinfo {editor} {\bibfnamefont {S.-V.}\
  \bibnamefont {Stratila}}, \ and\ \bibinfo {editor} {\bibfnamefont {D.-V.}\
  \bibnamefont {Voiculescu}}}\ (\bibinfo  {publisher} {Springer},\ \bibinfo
  {year} {1985})\ pp.\ \bibinfo {pages} {556--588}\BibitemShut {NoStop}%
\bibitem [{\citenamefont {Horn}\ and\ \citenamefont
  {Johnson}(1990)}]{Horn1990}%
  \BibitemOpen
  \bibfield  {author} {\bibinfo {author} {\bibfnamefont {R.~A.}\ \bibnamefont
  {Horn}}\ and\ \bibinfo {author} {\bibfnamefont {C.~R.}\ \bibnamefont
  {Johnson}},\ }\href@noop {} {\emph {\bibinfo {title} {Matrix Analysis}}}\
  (\bibinfo  {publisher} {Cambridge},\ \bibinfo {address} {UK},\ \bibinfo
  {year} {1990})\BibitemShut {NoStop}%
\bibitem [{\citenamefont {Neu}\ and\ \citenamefont {Speicher}(1994)}]{Neu1994}%
  \BibitemOpen
  \bibfield  {author} {\bibinfo {author} {\bibfnamefont {P.}~\bibnamefont
  {Neu}}\ and\ \bibinfo {author} {\bibfnamefont {R.}~\bibnamefont {Speicher}},\
  }\href {\doibase 10.1007/BF01316850} {\bibfield  {journal} {\bibinfo
  {journal} {Z. Phys. B}\ }\textbf {\bibinfo {volume} {95}},\ \bibinfo {pages}
  {101} (\bibinfo {year} {1994})}\BibitemShut {NoStop}%
\bibitem [{\citenamefont {Neu}\ and\ \citenamefont
  {Speicher}(1995{\natexlab{a}})}]{Neu1995a}%
  \BibitemOpen
  \bibfield  {author} {\bibinfo {author} {\bibfnamefont {P.}~\bibnamefont
  {Neu}}\ and\ \bibinfo {author} {\bibfnamefont {R.}~\bibnamefont {Speicher}},\
  }\href {http://iopscience.iop.org/0305-4470/28/3/004} {\bibfield  {journal}
  {\bibinfo  {journal} {J. Phys. A}\ }\textbf {\bibinfo {volume} {79}},\
  \bibinfo {pages} {L79} (\bibinfo {year} {1995}{\natexlab{a}})}\BibitemShut
  {NoStop}%
\bibitem [{\citenamefont {Neu}\ and\ \citenamefont
  {Speicher}(1995{\natexlab{b}})}]{Neu1995b}%
  \BibitemOpen
  \bibfield  {author} {\bibinfo {author} {\bibfnamefont {P.}~\bibnamefont
  {Neu}}\ and\ \bibinfo {author} {\bibfnamefont {R.}~\bibnamefont {Speicher}},\
  }\href {\doibase 10.1007/BF02179871} {\bibfield  {journal} {\bibinfo
  {journal} {J. Stat. Phys.}\ }\textbf {\bibinfo {volume} {80}},\ \bibinfo
  {pages} {1279} (\bibinfo {year} {1995}{\natexlab{b}})}\BibitemShut {NoStop}%
\bibitem [{\citenamefont {Blackman}\ \emph {et~al.}(1971)\citenamefont
  {Blackman}, \citenamefont {Esterling},\ and\ \citenamefont
  {Berk}}]{Blackman1971}%
  \BibitemOpen
  \bibfield  {author} {\bibinfo {author} {\bibfnamefont {J.}~\bibnamefont
  {Blackman}}, \bibinfo {author} {\bibfnamefont {D.}~\bibnamefont {Esterling}},
  \ and\ \bibinfo {author} {\bibfnamefont {N.}~\bibnamefont {Berk}},\ }\href
  {\doibase 10.1103/PhysRevB.4.2412} {\bibfield  {journal} {\bibinfo  {journal}
  {Phys. Rev. B}\ }\textbf {\bibinfo {volume} {4}},\ \bibinfo {pages} {2412}
  (\bibinfo {year} {1971})}\BibitemShut {NoStop}%
\bibitem [{\citenamefont {Movassagh}\ and\ \citenamefont
  {Edelman}(2011)}]{Movassagh2011a}%
  \BibitemOpen
  \bibfield  {author} {\bibinfo {author} {\bibfnamefont {R.}~\bibnamefont
  {Movassagh}}\ and\ \bibinfo {author} {\bibfnamefont {A.}~\bibnamefont
  {Edelman}},\ }\href {\doibase 10.1103/PhysRevLett.107.097205} {\bibfield
  {journal} {\bibinfo  {journal} {Phys. Rev. Lett.}\ }\textbf {\bibinfo
  {volume} {107}},\ \bibinfo {pages} {097205} (\bibinfo {year}
  {2011})}\BibitemShut {NoStop}%
\end{thebibliography}%

\end{document}